\documentclass[aps,prl,twocolumn,superscriptaddress]{revtex4-2}
\pdfoutput=1
\usepackage{graphicx}
\usepackage{amsmath}
\usepackage{amssymb}
\usepackage{siunitx}
\usepackage[version=4]{mhchem}

\begin{document}

\righthyphenmin=4
\lefthyphenmin=4

\title{Spin-orbit-enhanced magnetic surface second-harmonic generation in Sr$_2$IrO$_4$}

\author{K. L. Seyler}
\affiliation{
 Department of Physics, California Institute of Technology, Pasadena, CA 91125, USA
}
\affiliation{
 Institute for Quantum Information and Matter, California Institute of Technology, Pasadena, CA 91125, USA
}

\author{A. de la Torre}
\affiliation{
 Department of Physics, California Institute of Technology, Pasadena, CA 91125, USA
}
\affiliation{
 Institute for Quantum Information and Matter, California Institute of Technology, Pasadena, CA 91125, USA
}

\author{Z. Porter}
\affiliation{
 Materials Department, University of California, Santa Barbara, California 93106, USA
}

\author{E. Zoghlin}
\affiliation{
 Materials Department, University of California, Santa Barbara, California 93106, USA
}

\author{R. Polski}
\affiliation{
 Institute for Quantum Information and Matter, California Institute of Technology, Pasadena, CA 91125, USA
}
\affiliation{
 T. J. Watson Laboratory of Applied Physics, California Institute of Technology, Pasadena, CA 91125, USA
}

\author{M. Nguyen}
\affiliation{
 Department of Physics, California Institute of Technology, Pasadena, CA 91125, USA
}
\affiliation{
 Institute for Quantum Information and Matter, California Institute of Technology, Pasadena, CA 91125, USA
}

\author{S. Nadj-Perge}
\affiliation{
 Institute for Quantum Information and Matter, California Institute of Technology, Pasadena, CA 91125, USA
}
\affiliation{
 T. J. Watson Laboratory of Applied Physics, California Institute of Technology, Pasadena, CA 91125, USA
}

\author{S. D. Wilson}
\affiliation{
 Materials Department, University of California, Santa Barbara, California 93106, USA
}

\author{D. Hsieh}
\affiliation{
 Department of Physics, California Institute of Technology, Pasadena, CA 91125, USA
}
\affiliation{
 Institute for Quantum Information and Matter, California Institute of Technology, Pasadena, CA 91125, USA
}

\date{\today}

\begin{abstract}

An anomalous optical second-harmonic generation (SHG) signal was previously reported in Sr$_2$IrO$_4$ and attributed to a hidden odd-parity bulk magnetic state. Here we investigate the origin of this SHG signal using a combination of bulk magnetic susceptibility, magnetic-field-dependent SHG rotational anisotropy, and overlapping wide-field SHG imaging and atomic force microscopy measurements. We find that the anomalous SHG signal exhibits a two-fold rotational symmetry as a function of in-plane magnetic field orientation that is associated with a crystallographic distortion. We also show a change in SHG signal across step edges that tracks the bulk antiferromagnetic stacking pattern. While we do not rule out the existence of hidden order in Sr$_2$IrO$_4$, our results altogether show that the anomalous SHG signal in parent Sr$_2$IrO$_4$ originates instead from a surface-magnetization-induced electric-dipole process that is enhanced by strong spin-orbit coupling.

\end{abstract}

\maketitle

The layered square-lattice iridate \ce{Sr2IrO4} has earned recognition for its spin-orbit Mott state \cite{Kim2009-un} and various analogies to cuprate physics, including the emergence of a pseudogap \cite{Kim2014-cy,De_la_Torre2015-ki,Cao2016-ta} that evolves into a \textit{d}-wave gap at lower temperatures \cite{Kim2015-jb,Yan2015-lu}. Recently, optical second-harmonic generation (SHG) experiments reported evidence of unexpected broken spatial symmetries in \ce{Sr2IrO4} \cite{Zhao2015-lg}, suggestive of a hidden order. However, both the existence and the microscopic origin of a hidden order in \ce{Sr2IrO4} remain under intensive experimental \cite{Jeong2017-ul,Tan2020-nh,Murayama2020-cc} and theoretical investigation \cite{Lovesey2014-cw,Di_Matteo2016-eh,Chatterjee2017-bi,Zhou2017,Sumita2017-yo,Gao2018-hf}.

\begin{figure*}[!htb]
\includegraphics{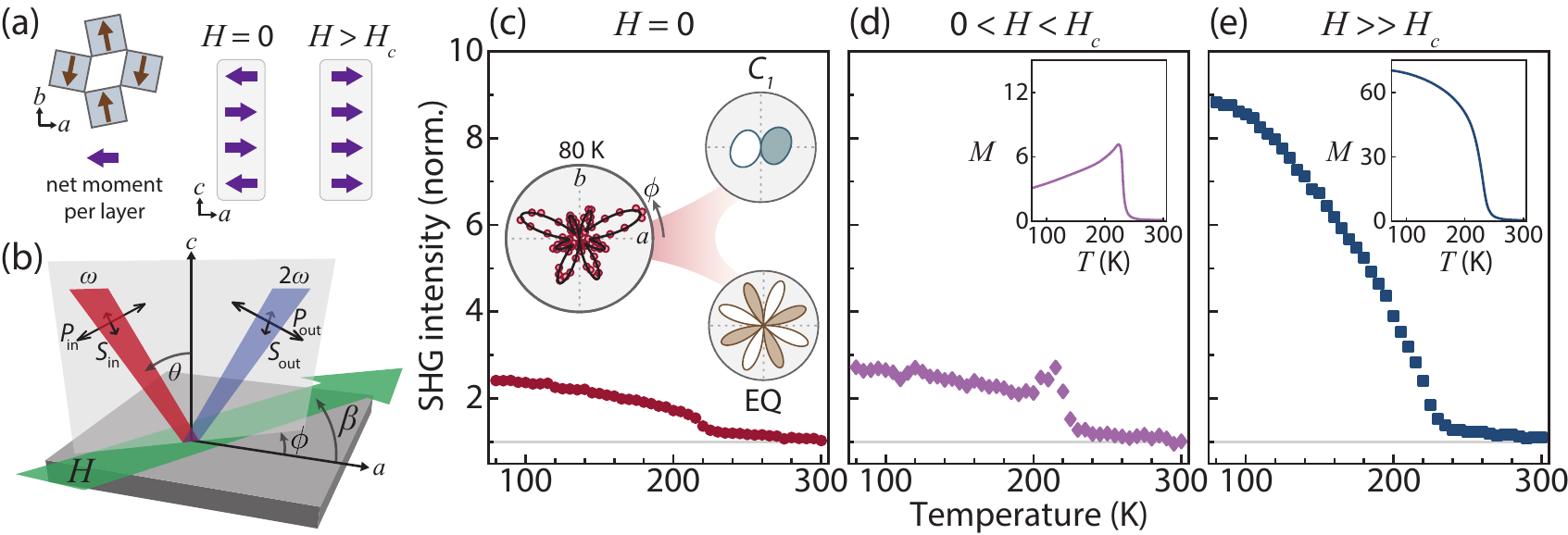}
\caption{\label{fig:m1} (a) Intralayer N\'eel order of the $J_{\textrm{eff}} = 1/2$ pseudospins and the $c$-axis stacking pattern of the net in-plane moments (purple arrows) at $H = 0$ (${-++-}$) and $H > H_c$ (${++++}$) below $T_N$. (b) Schematic of RA-SHG setup. The angle of incidence ($\theta$) is fixed to 10$^{\circ}$ while the scattering plane angle ($\phi$) varied. The fundamental input ($\omega$) and SH output ($2\omega$) beams can be selected as \textit{P}- or \textit{S}-polarized. An in-plane magnetic field $H$ (green arrow) can be applied at varying angles $\beta$ relative to [100]. Temperature dependence of \textit{P}\textsubscript{in}-\textit{S}\textsubscript{out} SHG intensities acquired in a field-cooled warming scheme for (c) $H = \SI{0}{\milli\tesla}$, (d) \SI{75}{\milli\tesla}, and (e) \SI{370}{\milli\tesla}, with $\phi$ fixed at the lobe of maximum intensity. The curves are normalized to the room-temperature value of the zero field data. Insets: (c) RA data for \textit{P}\textsubscript{in}-\textit{S}\textsubscript{out} polarization geometry measured at $T = \SI{80}{\kelvin}$ and $H = \SI{0}{\tesla}$. As reported in Ref.~\cite{Zhao2015-lg}, this pattern is produced by the interference between a $C_4$ crystallographic EQ term with a $C_1$ magnetic ED term as illustrated on the right. Filled and white lobes denote opposite optical phase. The relative lobe intensities are highly sensitive to small changes in the EQ and ED tensor elements, which are found to vary slightly between this work and Ref.~\cite{Zhao2015-lg}. (d) Magnetization curves for \SI{30}{\milli\tesla} and (e) \SI{370}{\milli\tesla} in units of $10^{-3} \mu_{\textrm{B}}$/Ir.}
\end{figure*}

The key feature of the SHG data in \ce{Sr2IrO4} is the onset of a new radiation process below the N\'eel temperature ($T_N \sim 230$ K) that lowers the rotational symmetry about the $c$ axis from $C_4$ to $C_1$. This is incompatible with its reported N\'eel structure \cite{Kim2009-un,Dhital2013-jj,Ye2013-pw,Boseggia2013-kd}, in which a canting-induced net ferromagnetic moment in each layer is stacked along the $c$ axis in a ${-++-}$ order so as to preserve $C_2$ symmetry [Fig.~\ref{fig:m1}(a)]. Rather, it was found to be consistent with electric-dipole (ED) radiation from a noncentrosymmetric $2'/m$ or $m1'$ magnetic point group \cite{Zhao2015-lg}, which bears the symmetries of a magnetoelectric loop-current phase \cite{Varma1997-oa}. However, alternative explanations were subsequently proposed \cite{Di_Matteo2016-eh}, including laser-induced rearrangement of the magnetic stacking and enhanced sensitivity to surface rather than bulk magnetic order.

Studying the field dependence of the anomalous $C_1$ SHG signal can help reveal its origin. Previous works showed that applying an in-plane magnetic field $H > H_c$ (\SI{\sim 200}{\milli\tesla} for temperatures well below $T_N$) induces a metamagnetic transition to a $C_2$-breaking ${++++}$ state [Fig.~\ref{fig:m1}(a)], whose net ferromagnetic moment rotates with the in-plane field orientation \cite{Porras2019-qb}. In contrast, the proposed magnetoelectric order parameter can only couple linearly to $H$ in the presence of an additional electric field \cite{Shekhter2009-zm}. Here we report field-dependent SHG rotational anisotropy (RA), magnetic susceptibility, as well as overlapping atomic force and SHG microscopy measurements on parent \ce{Sr2IrO4}. While we do not rule out the presence of hidden order in \ce{Sr2IrO4} \cite{Jeong2017-ul,Tan2020-nh,Murayama2020-cc}, our data show that the $C_1$ SHG signal arises from an unexpectedly strong surface-magnetization-induced ED process in the ${-++-}$ state, rather than from laser-induced or intrinsic hidden orders.

Our experiments were performed on cleaved single crystals of \ce{Sr2IrO4} \cite{SI}.\nocite{Chen2018-fh,Sung2016-av,Harter2015-iy,Birss1964-bk,Fiebig2005} The geometry of our magnetic-field-tunable RA-SHG setup is shown in Fig.~\ref{fig:m1}(b). In zero field, the low temperature RA ($\phi$-dependence) pattern is produced from the interference between a $C_4$ crystallographic electric-quadrupole (EQ) process and the anomalous $C_1$ process [Fig.~\ref{fig:m1}(c)]. By tracking the intensity of the strongest lobe versus temperature, we confirm an order-parameter-like onset of the latter near $T = \SI{230}{\kelvin}$ in agreement with previous work \cite{Zhao2015-lg}. In the presence of a weak in-plane field ($H < H_c$), no change is observed in the high-temperature EQ intensity, but there is slight enhancement of the low-temperature SHG intensity along with the emergence of a peak structure just below $T_N$ [Fig.~\ref{fig:m1}(d)]. This is reminiscent of bulk magnetization data acquired under similar field conditions, which exhibits a ferromagnetic upturn at $T_N$ followed by a sharp drop as the ${-++-}$ state is stabilized [inset Fig.~\ref{fig:m1}(d)]. A closer comparison of their temperature dependence suggests that the low-field SHG curve is a superposition of the zero-field curve and a new contribution that tracks the bulk magnetization. This is further corroborated by the high-field ($H > H_c$) SHG data [Fig.~\ref{fig:m1}(e)] that shows a large intensity increase relative to the low-field case and a disappearance of the peak structure, mirroring the behavior of the bulk magnetization in the ${++++}$ state [inset Fig.~\ref{fig:m1}(e)].

The data in Fig.~\ref{fig:m1} reveal three contributions to the field-dependent SHG response: a crystallographic EQ term and an anomalous $C_1$ term that both persist in zero field, plus a bulk magnetization-dependent term that is strongly enhanced in the ${++++}$ state. This eliminates the possibility of the sample already being in a laser-induced ${++++}$ state at zero field. To rule out the possibility of alternative laser-induced stacking patterns \cite{Di_Matteo2016-eh}, we repeated our zero-field measurements on a fresh sample using two orders of magnitude weaker laser fluence and observed the same trend as in Fig.~\ref{fig:m1}(c) \cite{SI}. Furthermore, we performed time-resolved SHG experiments and observed a strong suppression of the SHG intensity after a pump pulse, rather than an enhancement \cite{SI}. These tests show that the $C_1$ term arises from an intrinsic rather than laser-induced effect.

\begin{figure*}[!htb]
\includegraphics{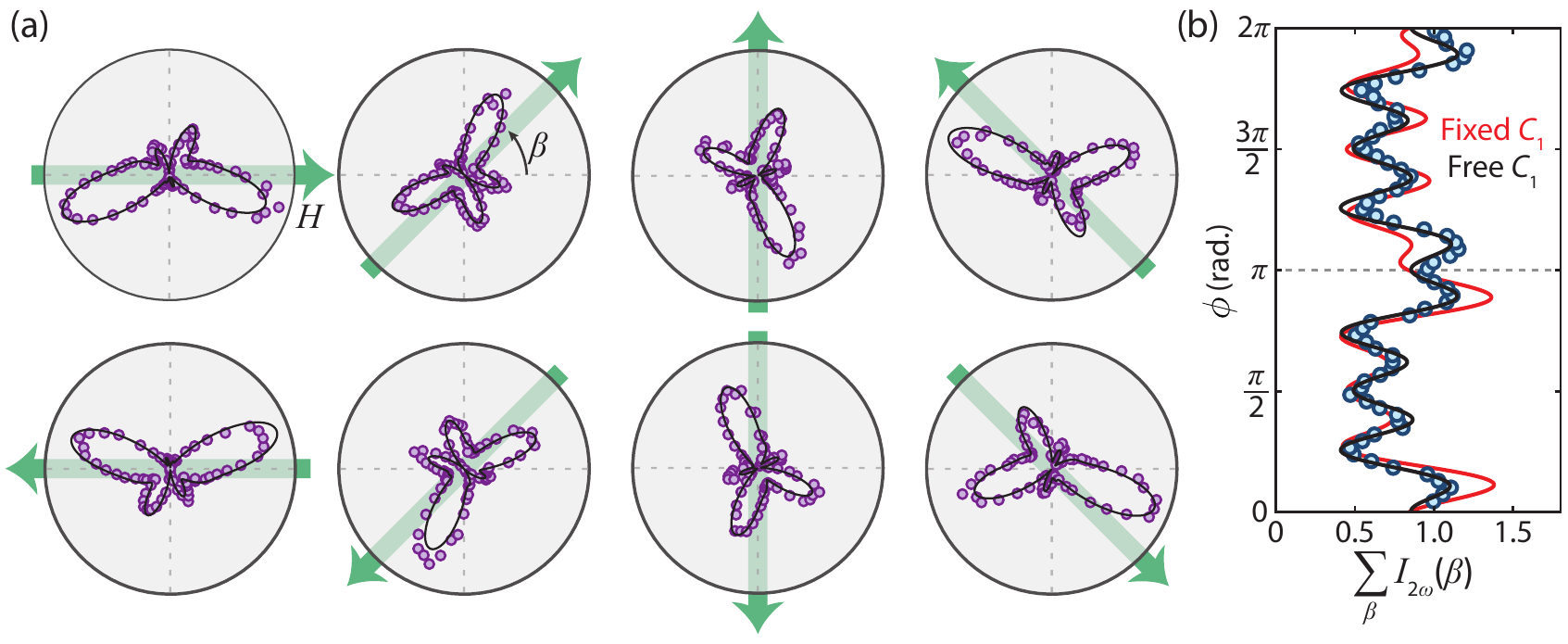}
\caption{\label{fig:m2} (a) RA-SHG patterns acquired in \textit{P}\textsubscript{in}-\textit{S}\textsubscript{out} geometry at \SI{80}{\kelvin} for different applied magnetic field angles ($\beta$) with a fixed field strength of \SI{370}{\milli\tesla}. Solid black lines are fits to the bulk EQ $+$ MD model described in the main text using a fixed set of susceptibility tensor elements and only varying $\beta$ \cite{SI}. (b) Summation of the field-dependent RA data over all $\beta$, where $\beta$ runs from \ang{0} to \ang{345} in \ang{15} increments. Solid lines show fits to a model that includes both the bulk EQ $+$ MD terms and the anomalous $C_1$ term, where the latter either rotates with $\beta$ (black) or is independent of $\beta$ (red).
}
\end{figure*}

To examine how the anomalous $C_1$ term varies with magnetic field orientation, we collected RA-SHG patterns as a function of the in-plane field angle ($\beta$) with $|H|$ fixed at 370 mT. Figure~\ref{fig:m2}(a) shows that the RA patterns continue to exhibit $C_1$ symmetry in the ${++++}$ state and rotate by \ang{180} upon reversal of the field direction ($H \rightarrow -H$). This implies that both the bulk magnetization-dependent term and the anomalous $C_1$ term couple linearly to a magnetic field. By summing the RA-SHG data over all $\beta$ to isolate the $\beta$-independent symmetries [Fig.~\ref{fig:m2}(b)], we confirm that no $C_1$ component remains, which is evidence against the magnetoelectric order parameter interpretation of the SHG response \cite{SI}. Surprisingly, despite the reported tetragonal symmetry of the lattice, the summed data exposes a twofold rotational symmetry, indicating an inequivalent response to fields along $[100]$ and $[010]$. In \ce{Sr2IrO4}, coupling of the pseudospins to an orthorhombically deformed lattice below $T_N$ naturally induces uniaxial anisotropy \cite{Liu2019-qe,Porras2019-qb}, which can lead to a twofold symmetry in $\beta$ when $H$ is below or near $H_c$ \cite{Wang2014-ql,Wang2019-oo}. However, our SHG experiments were performed in a high-field regime where the bulk magnetization is constant and rotates rigidly with the field direction \cite{Porras2019-qb,SI}. These observations were reproduced across multiple samples and up to temperatures as high as \SI{220}{\kelvin}, where pseudospin-lattice coupling plays a smaller role \cite{Porras2019-qb,SI}. This points to an $a$-$b$ symmetry breaking distortion already existing above $T_N$.

The nonlinear susceptibility tensors governing the bulk magnetization-dependent and anomalous SHG processes must both be $c$-type (time-reversal odd) by virtue of their observed linear coupling to $H$. Since the centrosymmetric $2'/m'$ magnetic point group of the ${++++}$ state forbids ED SHG, it was proposed \cite{Di_Matteo2016-eh} that the former can be associated with a bulk magnetic-dipole (MD) process of the type
\begin{equation}
\label{eqn:m1}
P_i(2\omega) \propto \chi_{ijk}^{\textrm{MD} (c)}(\textbf{M})E_j(\omega)H_k(\omega),
\end{equation}
where an incident electromagnetic field with frequency $\omega$ induces an oscillatory polarization at $2\omega$. To fit our high-field data [Fig.~\ref{fig:m2}(a)], we express the axial $c$ tensor $\chi_{ijk}^{\textrm{MD} (c)}(\textbf{M})$ as $\chi_{ijkl}^{\textrm{MD} (i)}M_l$, where $\chi_{ijkl}^{\textrm{MD} (i)}$ is a polar $i$ tensor (time-reversal even) that respects the high temperature crystallographic point group and $\textbf{M}=(M\cos\beta,M\sin\beta,0)$ is the static magnetization of the sample \cite{SI}. This term is then coherently added to a constant crystallographic EQ term. A good simultaneous fit to the entire set of field-dependent RA patterns [Fig.~\ref{fig:m2}(a)], where the $\chi_{ijkl}^{\textrm{MD} (i)}$ elements are fixed and only $\beta$ varies, can be achieved only by assuming that $\chi_{ijkl}^{\textrm{MD} (i)}$ breaks tetragonal symmetry \cite{SI}. This reflects our observation that the magnetic field response along $a$ and $b$ differ.

Attributing the N\'eel-order-induced contribution to a $c$-type MD process naturally explains why there is strong bulk SHG from the ${++++}$ state but not the ${-++-}$ state. Since the MD susceptibility of each layer is time-reversal odd, layers with opposite magnetization radiate \ang{180} out of phase, leading to overall destructive (constructive) interference for antiferromagnetic (ferromagnetic) stacking. It has been shown that ${-++-}$ stacking is stabilized by competing first and second nearest-neighbor interlayer exchange coupling \cite{Takayama2016}. Therefore, although we excluded bulk ${++++}$ stacking as the source of the zero-field SHG, one could posit that crystal termination forces a ${++++}$ stacking near the surface of a bulk ${-++-}$ ordered sample. In this scenario, we expect the zero-field RA patterns to map onto the high-field patterns by simply scaling $\chi_{ijk}^{\textrm{MD}(c)}(\textbf{M})$. However, our fitting suggests this is not the case \cite{SI}, leaving pure surface SHG as the remaining plausible candidate.

\begin{figure}[!htb]
\includegraphics{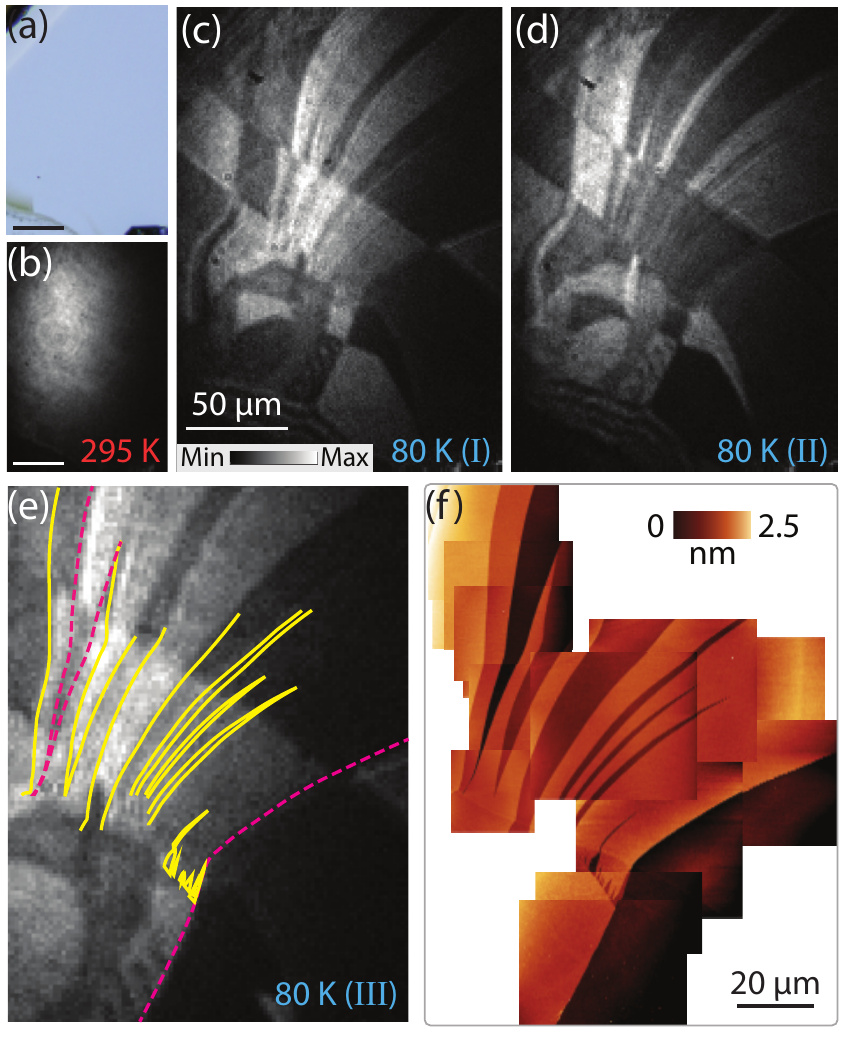}
\caption{\label{fig:m3} (a) White light microscopy and (b) wide-field SHG image of (001) cleaved \ce{Sr2IrO4} acquired at \SI{295}{\kelvin} in zero magnetic field. \textit{P}\textsubscript{in}-\textit{P}\textsubscript{out} geometry was used with ${\phi\sim\ang{185}}$. Scale bars, \SI{50}{\um}. (c) SHG image acquired after first and (d) second cool down to \SI{80}{\kelvin} starting from \SI{295}{\kelvin}. (e) Zoomed-in SHG image acquired after third cool down. (f) Composite image of several atomic force microscopy height maps. The yellow solid (magenta dashed) lines in (e) mark where a monolayer (bilayer) step was observed in the height maps.}
\end{figure}

Since inversion symmetry is naturally broken at the surface of \ce{Sr2IrO4}, a zero-field $C_1$ term can in principle arise from a surface-magnetization-induced $c$-type ED SHG process
\begin{equation}
\label{eqn:m2}
P_i(2\omega)\propto\chi_{\textrm{s},ijk}^{\textrm{ED}(c)}(\textbf{M}_\textrm{s})E_j(\omega)E_k(\omega),
\end{equation}
where $\chi_{\textrm{s},ijk}^{\textrm{ED}(c)}(\textbf{M}_\textrm{s})$ can be expressed in terms of the surface magnetization as $\chi_{\textrm{s},ijkl}^{\textrm{ED}(i)}M_{\textrm{s},l}$ \cite{SI}. To explore this possibility, we performed SHG microscopy in conjunction with atomic force microscopy on ultraflat $(001)$ surfaces of \ce{Sr2IrO4} prepared by cleaving. These surfaces appear optically pristine in standard white light microscopy as well as wide-field SHG imaging at room temperature [Figs.~\ref{fig:m3}(a) and (b)]. However, after zero-field cooling to \SI{80}{\kelvin}, a patchwork of bright and dark regions emerge in the SHG image [Fig.~\ref{fig:m3}(c)], with the length scale of the patches varying from \SIrange{\sim 10}{100}{\um}. The intensity variations reflect the presence of four $C_1$ orientations and our choice of scattering plane angle ($\phi\sim\ang{185}$). Following thermal cycles across $T_N$ (\SI{80}{\kelvin} $\rightarrow$ \SI{295}{\kelvin} $\rightarrow$ \SI{80}{\kelvin}), the patches switch between bright and dark but the boundaries do not change [Figs.~\ref{fig:m3}(d) and (e)]. A close inspection of previous SHG imaging data on \ce{Sr2IrO4} \cite{Zhao2015-lg}, being careful to account for the shifted illumination between images, shows fixed boundaries similar to our data, suggestive of pinning to structural features.

To search for direct correlations between the spatial distribution of SHG patches and structural features, we show a large area topographic survey of the same sample in Fig.~\ref{fig:m3}(f), which is a composite image of height maps assembled from several atomic force microscopy scans. The maps reveal an atomically smooth surface that is interrupted by long boundaries where the height changes by either a monolayer (\SI{\sim 0.65}{\nm}) or bilayer (\SI{\sim 1.3}{\nm}) \cite{Crawford1994-xh}. We overlay these boundaries atop the SHG image in Fig.~\ref{fig:m3}(e), with monolayer and bilayer steps marked by solid and dashed lines, respectively. We find that each topographical step correlates with the boundary of an SHG patch. But interestingly, the reverse is not true; there are clear patch boundaries that are not associated with topographical steps.

\begin{figure}[!htb]
\includegraphics{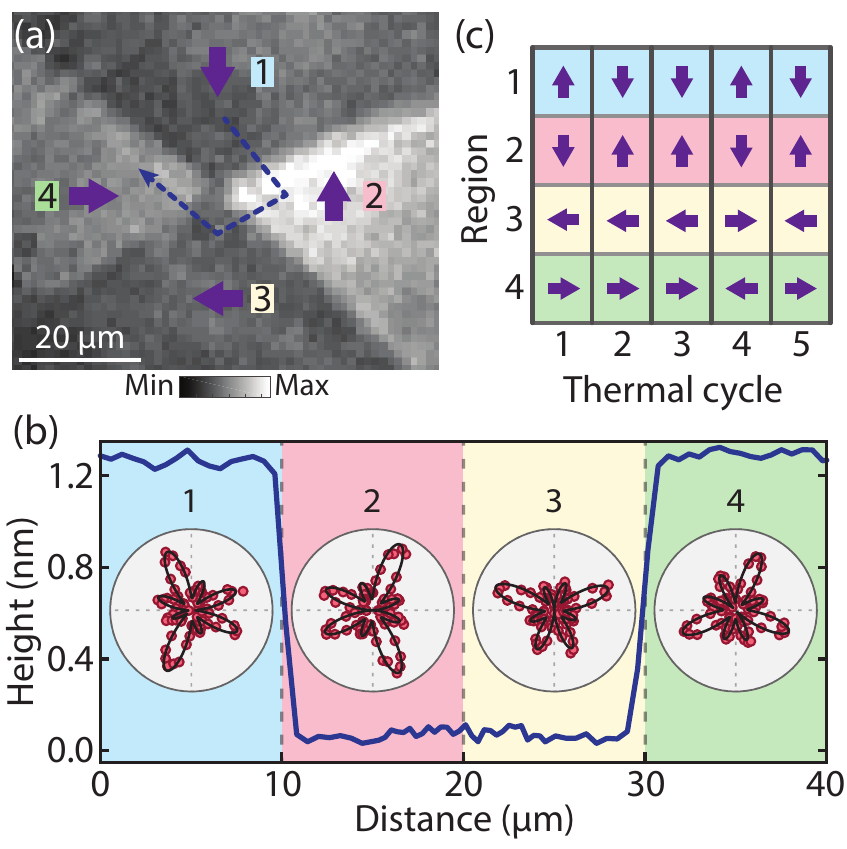}
\caption{\label{fig:m4} (a) Zoomed-in SHG image acquired in \textit{P}\textsubscript{in}-\textit{P}\textsubscript{out} geometry with $\phi\sim\ang{185}$ at 80 K and zero magnetic field. The local orientations of the $C_1$ order parameter (purple arrows) were determined by scanning RA measurements. (b) Atomic force microscopy height profile along a line cut shown by the dashed line in (a). Insets show the local \textit{P}\textsubscript{in}-\textit{S}\textsubscript{out} RA pattern in each region. Black lines are fits to the EQ + ED model described in the main text. (c) Order parameter orientation at \SI{80}{\kelvin} in each of the four regions after successive thermal cycles from \SI{80}{\kelvin} $\rightarrow$ \SI{295}{\kelvin} $\rightarrow$ \SI{80}{\kelvin}.}
\end{figure}

To understand these observations, we focus on regions where four distinct SHG patches intersect. Figure~\ref{fig:m4}(a) shows a representative example. By performing scanning RA measurements over this area, we find that the four patches correspond to four different $C_1$ order parameter orientations denoted by arrows. Across both the ${1 \rightarrow 4}$ and ${2 \rightarrow 3}$ boundaries, the surface is atomically flat with no step edges, but the order parameter undergoes a \ang{90} rotation [Fig.~\ref{fig:m4}(b)]. This is most likely caused by a crystallographic twin boundary \cite{SI} in which the $a$ and $b$ axes are interchanged. The presence of such twinning has been previously proposed to explain neutron diffraction results \cite{Dhital2013-jj,Ye2013-pw}. By measuring the magnetic-field-dependent RA response from each of the twin regions, we see evidence of $a$-$b$ symmetry breaking, as displayed in Fig.~\ref{fig:m2}(a), with axes switched between the regions \cite{SI}. Moreover, with repeated thermal cycling across $T_N$, each region is observed to randomly undergo \ang{0} or \ang{180}, but never \ang{90} or \ang{270}, order parameter rotations [Fig.~\ref{fig:m4}(c)], confirming this interpretation.

On the other hand, upon crossing from region ${1 \rightarrow 2}$ or from region ${3 \rightarrow 4}$, one traverses a bilayer step edge that coincides with a \ang{180} rotation of the order parameter [Fig.~\ref{fig:m4}(b)]. This is consistent with surface-magnetization-induced SHG from the established ${-++-}$ stacking since $\textbf{M}_\textrm{s}$ reverses every two layers. It also further rules out laser-induced ${++++}$ and ${-+-+}$ stacking that was previously proposed to explain the SHG data \cite{Di_Matteo2016-eh}. Upon successive thermal cycling across $T_N$, the order parameter orientations in regions 1 and 2 are always anti-correlated, as is the case for regions 3 and 4 [Fig.~\ref{fig:m4}(c)], whereas no correlation exists across the twin boundary. This indicates that regions 1 and 2 form part of a single magnetic domain with ${-++-}$ stacking, and that regions 3 and 4 form part of another single magnetic domain that is \ang{90} rotated. We note that a survey of all patches within the field of view shown in Figs.~\ref{fig:m3}(e) and \ref{fig:m3}(f) reveal that most of the terrace steps correlate with twin boundaries, suggesting that \ce{Sr2IrO4} crystals tend to fracture along them. We could not find any monolayer steps that were not associated with a twin boundary.

Surface-magnetization-induced ED SHG is typically weaker than its crystallographic counterpart. This is well documented for thin-film $3d$ transition metals such as Ni(110) \cite{Kirilyuk2005-lj} or Fe(110) \cite{Reif1991-dx}, for which the ratio ${\rho\equiv\chi^{\textrm{ED}(c)}_\textrm{s}(\textbf{M}_\textrm{s})/\chi^{\textrm{ED}(i)}_\textrm{s}(\textbf{M}_\textrm{s}{=}0)}$ is on the order of 0.1. The canted N\'{e}el structure of \ce{Sr2IrO4}, which is stabilized by a combination of conventional superexchange and Dzyaloshinskii-Moriya interactions \cite{Jackeli2009-kc}, results in a small $\textrm{M}_\textrm{s}$. Since the surface magnetization of \ce{Sr2IrO4} (${\sim}8 \times 10^{-2} \mu_{\textrm{B}}/\textrm{Ir}$) \cite{Chen2015-jv} is approximately an order of magnitude weaker than that of Ni or Fe films \cite{Freeman1987-ml}, one naively expects a much smaller value of $\rho$. Previous SHG experiments on \ce{Sr2IrO4} showed that the crystallographic contribution predominantly arises from a bulk $\chi^{\textrm{EQ}(i)}$ process rather than a $\chi^{\textrm{ED}(i)}_\textrm{s}$ process \cite{Torchinsky2015-ye}, in keeping with weak inversion symmetry breaking at the surface due to its quasi-2D structure. Therefore, the existence of an appreciable $\chi^{\textrm{ED}(c)}_\textrm{s}(\textbf{M}_\textrm{s})$ to $\chi^{\textrm{EQ}(i)}$ ratio is surprising and explains why it was discounted in Ref.~\cite{Zhao2015-lg}. To understand this phenomenon, we appeal to a perturbative calculation of surface-magnetization-induced SHG \cite{Pan1989}, which showed that $\rho \approx \textrm{M}_\textrm{s} A b$, where $A$ is the spin-orbit coupling for the free atom \cite{Goudsmit1928} and $b$ is a constant that depends on the band energies of the material. Since $A$ increases significantly going from 3$d$ to 5$d$ transition metals, it makes sense that \ce{Sr2IrO4} can still exhibit a large $\rho$ despite $\textrm{M}_\textrm{s}$ being small.

Recent polarized neutron diffraction \cite{Jeong2017-ul} and torque magnetometry \cite{Murayama2020-cc} measurements have suggested the presence of a bulk magnetoelectric loop-current order in \ce{Sr2IrO4}. There is also evidence that it survives and onsets above $T_N$ in Rh-doped samples \cite{Zhao2015-lg,Jeong2017-ul,Tan2020-nh,Murayama2020-cc}. However, these works point toward a magnetoelectric loop-current order that is ferroically stacked along the $c$ axis, which is incompatible with our observation of an SHG signal that switches sign every two layers. While our work does not rule out the existence of such a hidden order, it firmly establishes that the anomalous SHG signal in parent \ce{Sr2IrO4} is instead dominated by a process induced by surface magnetization. More generally, our results show that magnetization-induced SHG can be significantly enhanced in strongly spin-orbit coupled materials.

\begin{acknowledgments}
We acknowledge helpful conversations with Gang Cao, Nicholas Laurita, Sergio Di Matteo, Mike Norman, Alon Ron, Chandra Varma and Liuyan Zhao. This work is supported by an ARO PECASE award W911NF-17-1-0204. D.H. also acknowledges support for instrumentation from the David and Lucile Packard Foundation and from the Institute for Quantum Information and Matter (IQIM), an NSF Physics Frontiers Center (PHY-1733907). K.L.S. acknowledges a Caltech Prize Postdoctoral Fellowship. A.d.l.T. acknowledges support from the Swiss National Science Foundation through an Early Postdoc Mobility Fellowship (P2GEP2$\_$165044). S.D.W., E.Z. and Z.P. acknowledge support from ARO Award No. W911NF-16-1-0361. R.P. acknowledges support from IQIM.
\end{acknowledgments}

\providecommand{\noopsort}[1]{}\providecommand{\singleletter}[1]{#1}%

\newpage
\pagenumbering{gobble}
\begin{figure}
   \vspace*{-2cm}
   \hspace*{-2cm}
    \centering
    \includegraphics[page=1]{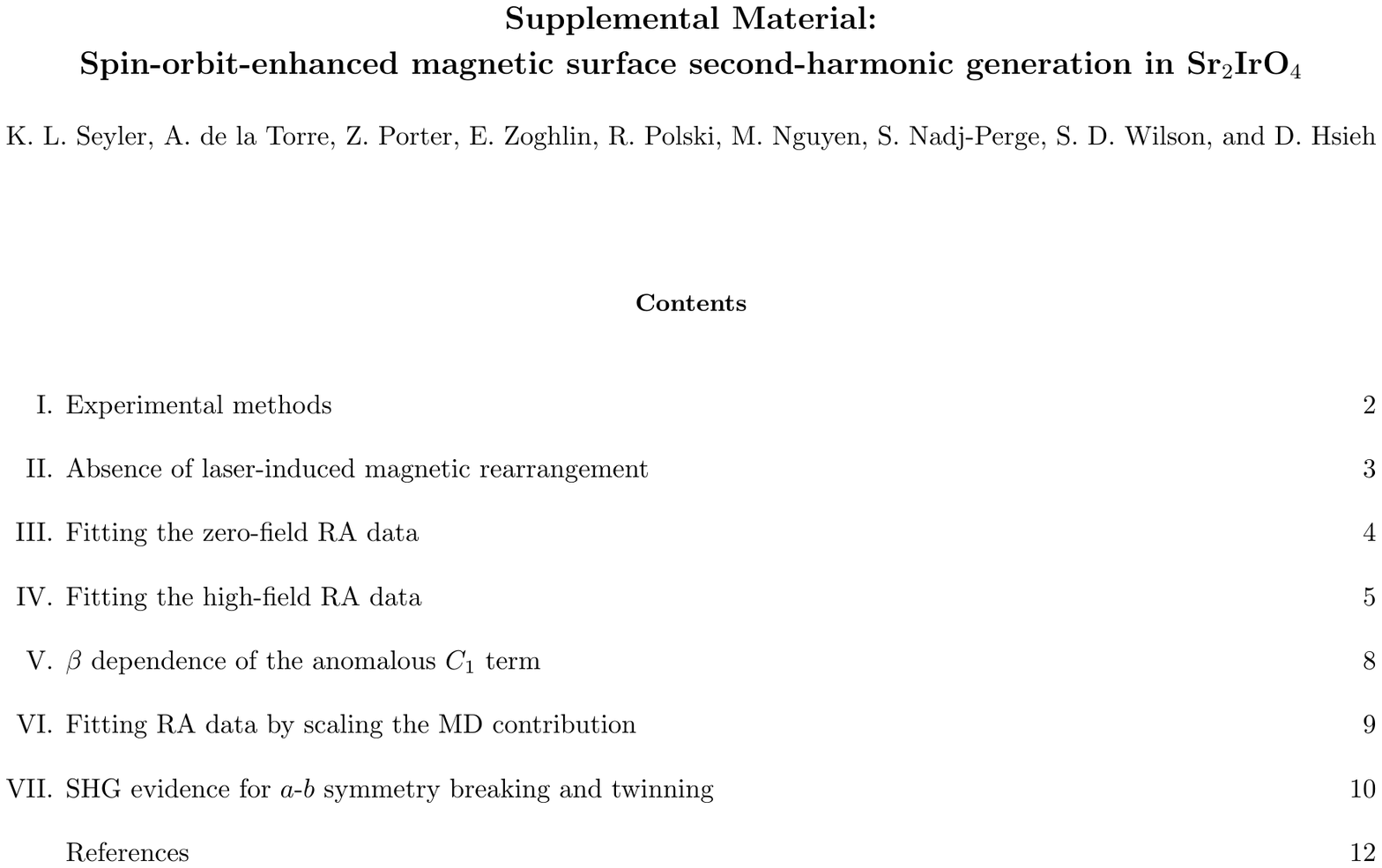}
    \caption{Caption}
    \label{fig:my_label}
\end{figure}
\begin{figure}
   \vspace*{-2cm}
   \hspace*{-2cm}
    \centering
    \includegraphics[page=2]{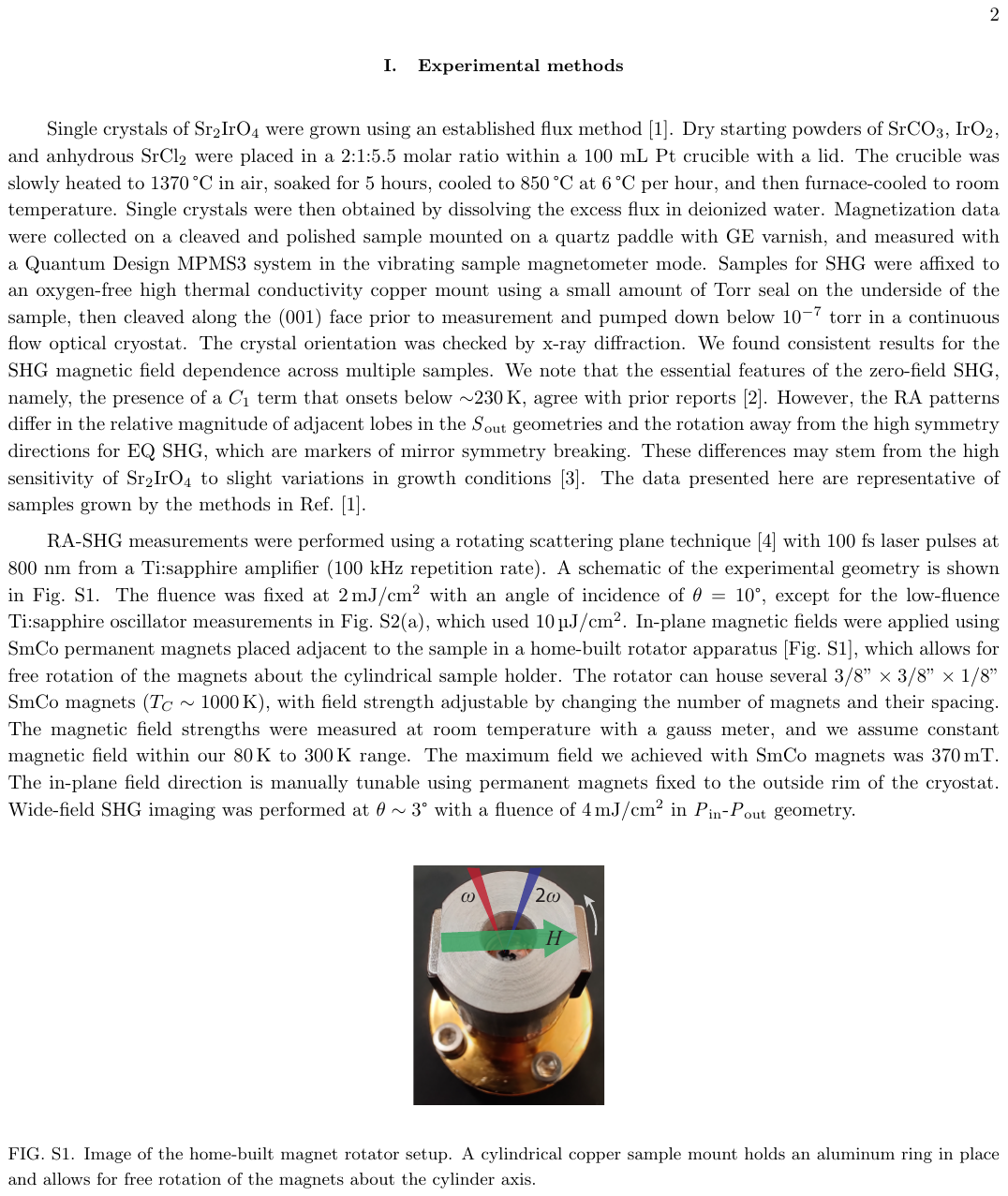}
    \caption{Caption}
    \label{fig:my_label}
\end{figure}
\begin{figure}
   \vspace*{-2cm}
   \hspace*{-2cm}
    \centering
    \includegraphics[page=3]{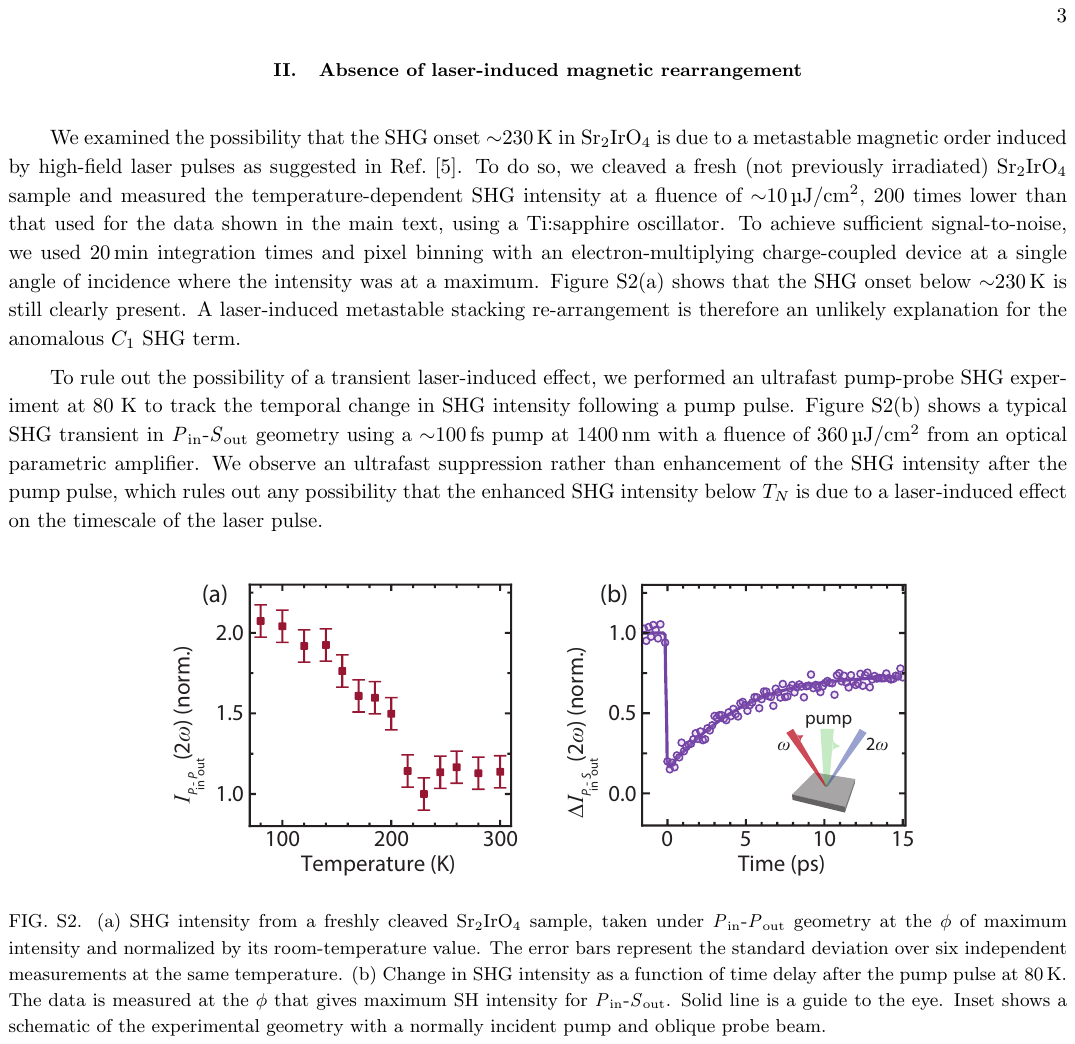}
    \caption{Caption}
    \label{fig:my_label}
\end{figure}
\begin{figure}
   \vspace*{-2cm}
   \hspace*{-2cm}
    \centering
    \includegraphics[page=4]{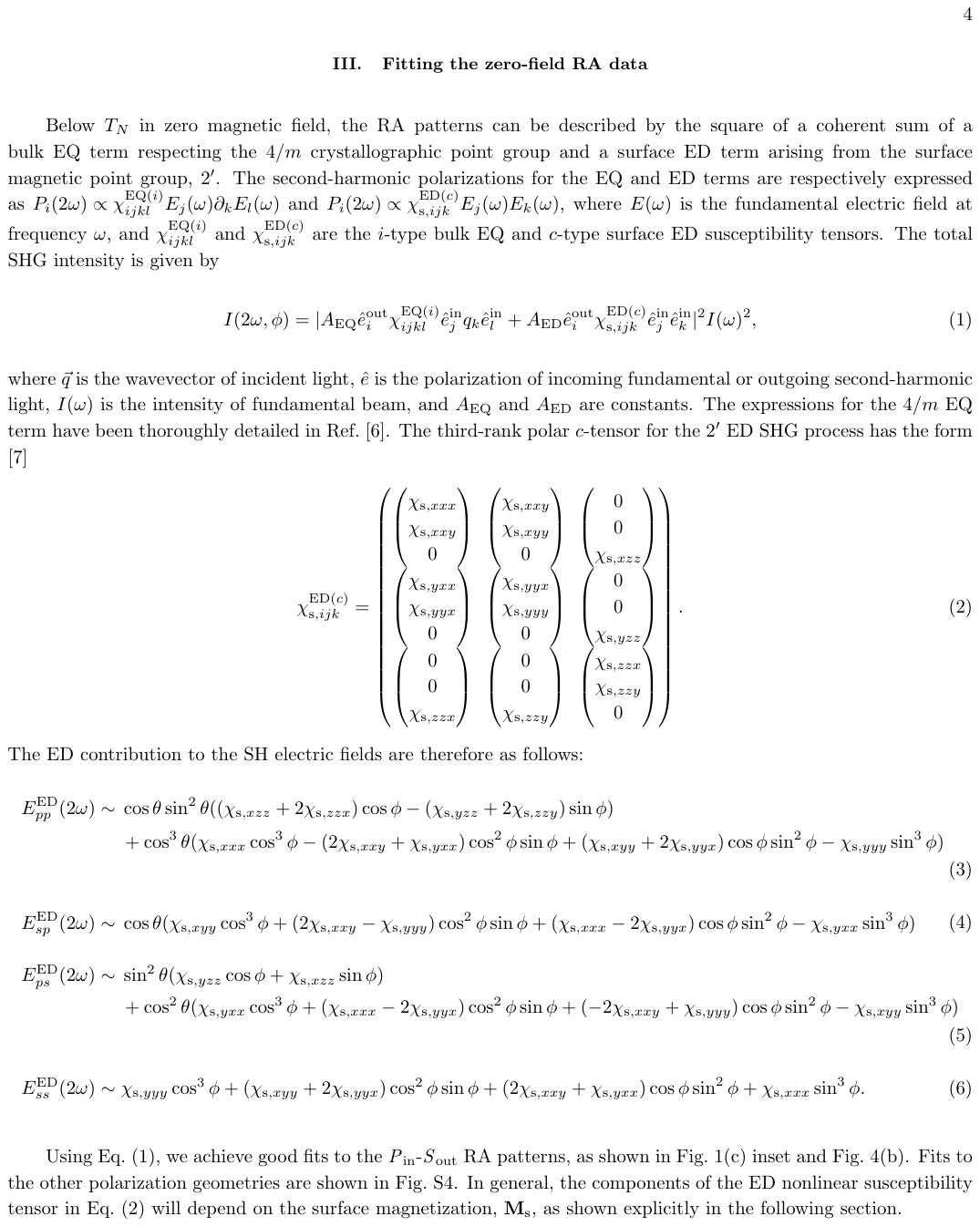}
    \caption{Caption}
    \label{fig:my_label}
\end{figure}
\begin{figure}
   \vspace*{-2cm}
   \hspace*{-2cm}
    \centering
    \includegraphics[page=5]{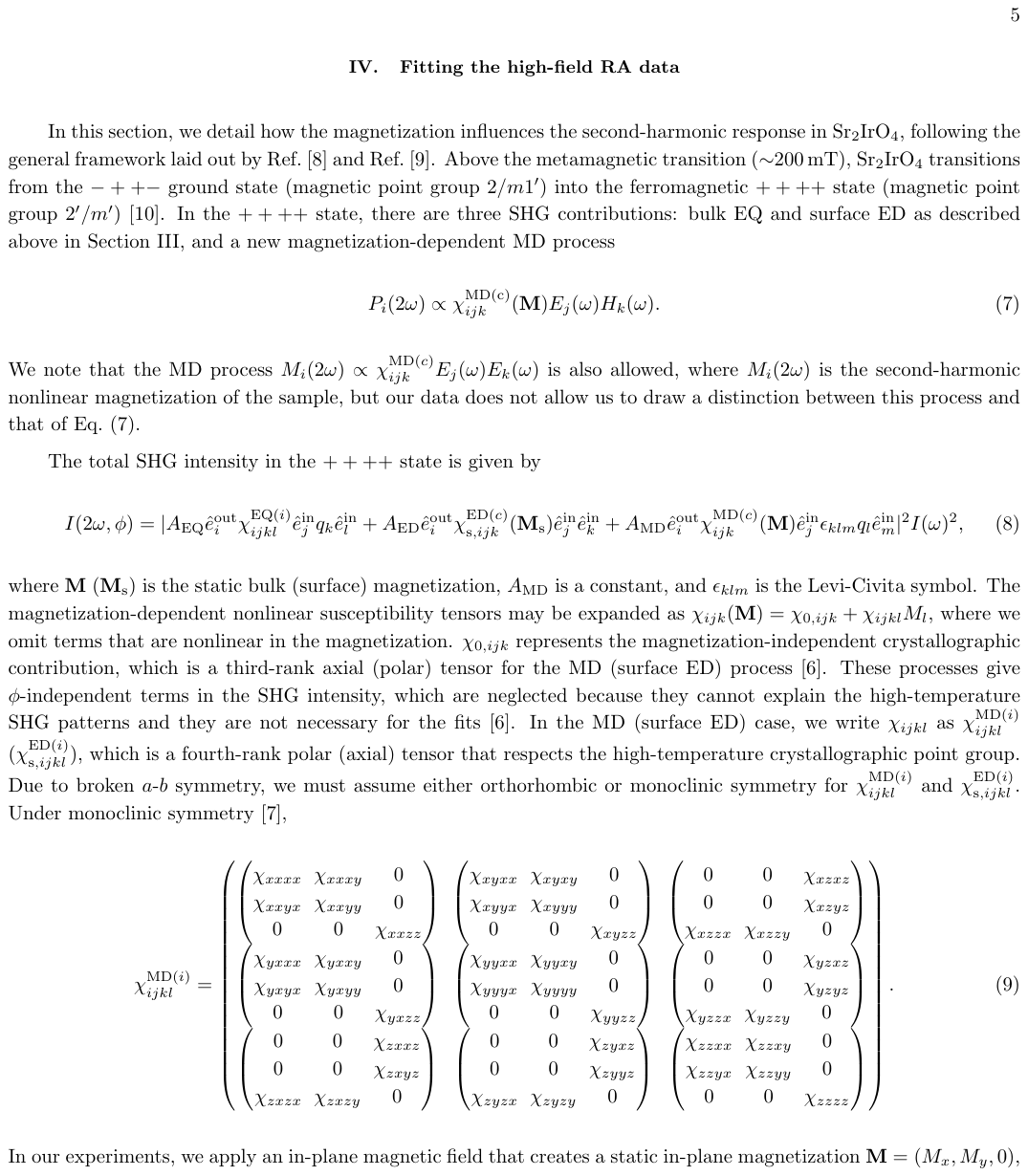}
    \caption{Caption}
    \label{fig:my_label}
\end{figure}
\begin{figure}
   \vspace*{-2cm}
   \hspace*{-2cm}
    \centering
    \includegraphics[page=6]{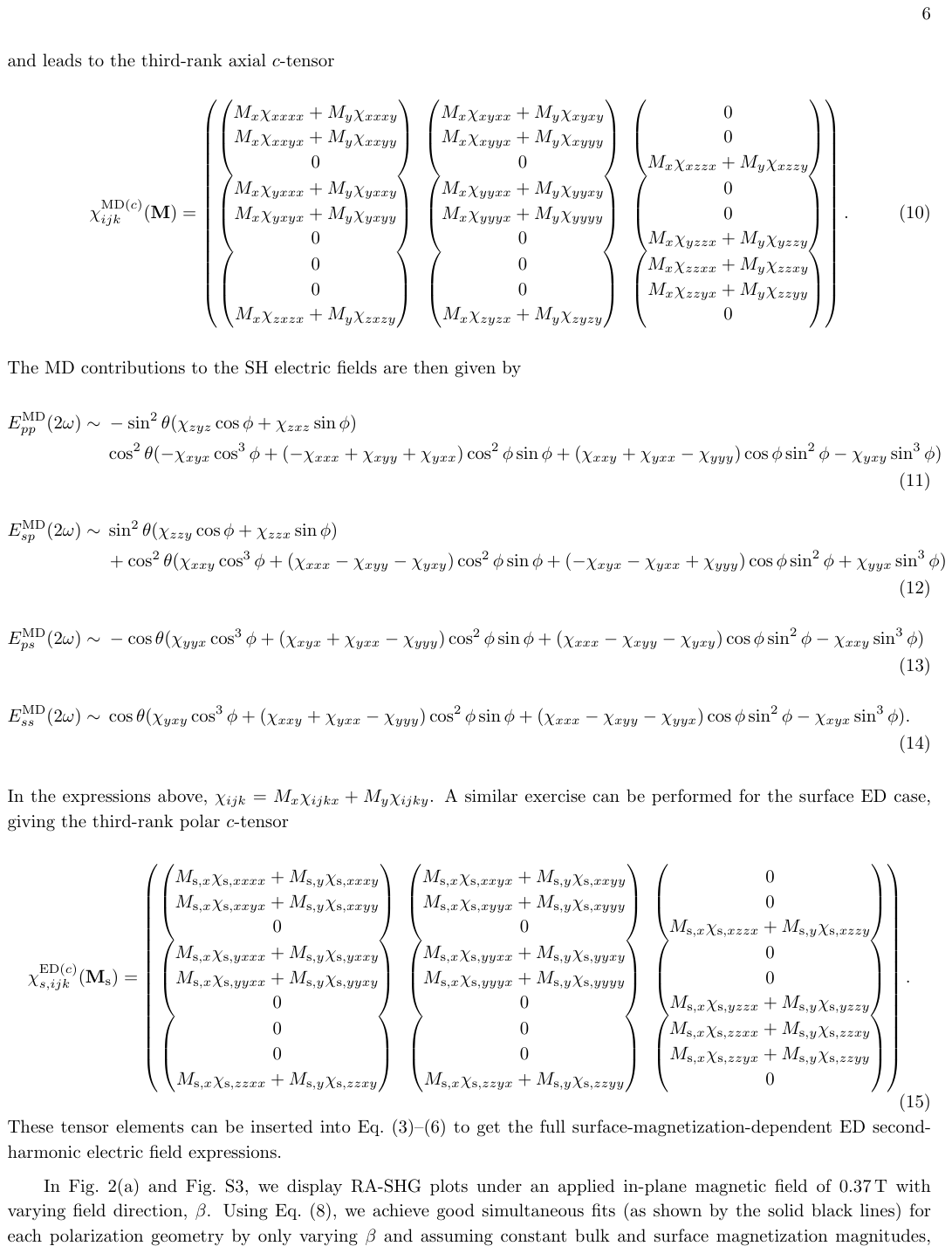}
    \caption{Caption}
    \label{fig:my_label}
\end{figure}
\begin{figure}
   \vspace*{-2cm}
   \hspace*{-2cm}
    \centering
    \includegraphics[page=7]{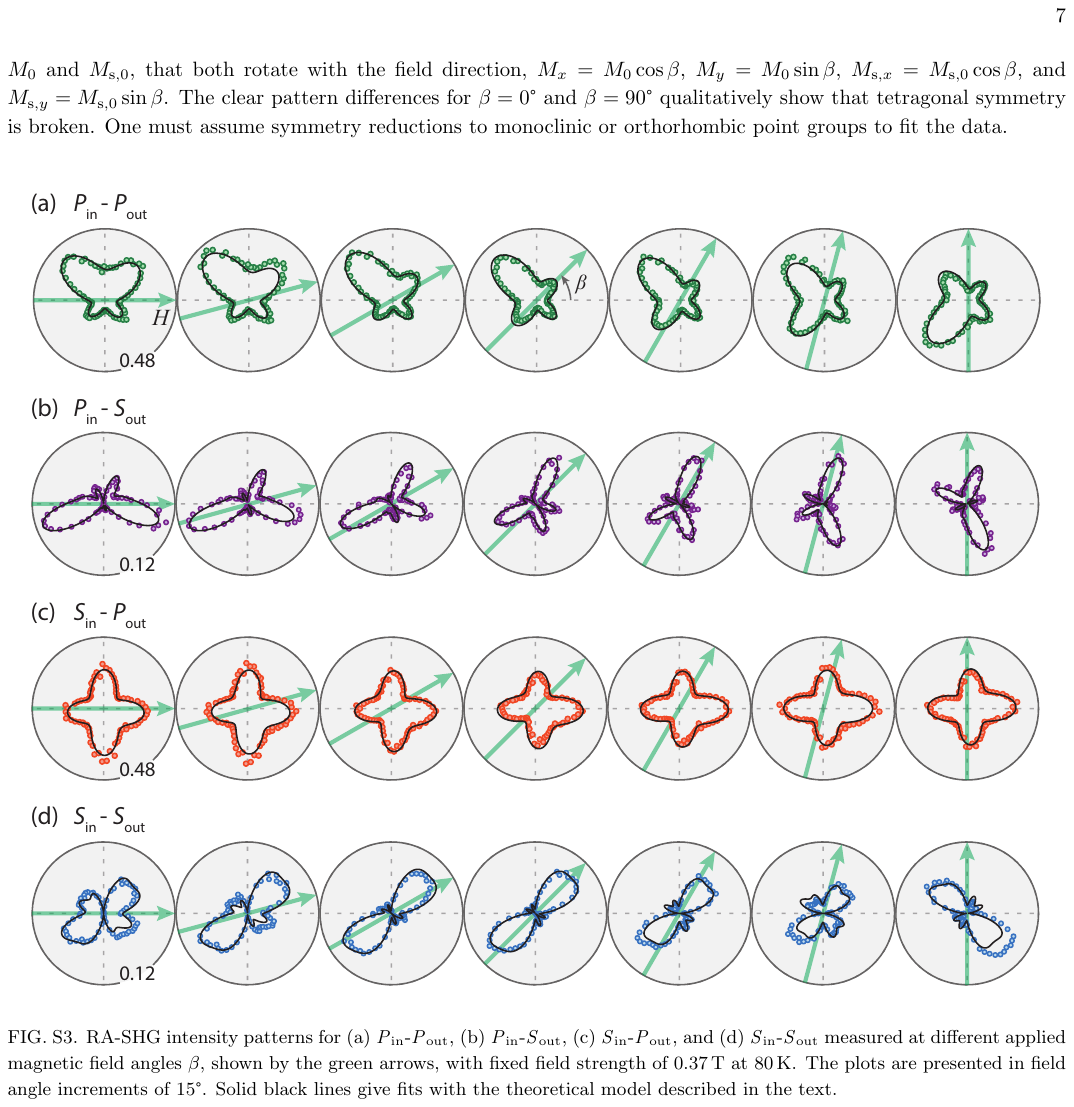}
    \caption{Caption}
    \label{fig:my_label}
\end{figure}
\begin{figure}
   \vspace*{-2cm}
   \hspace*{-2cm}
    \centering
    \includegraphics[page=8]{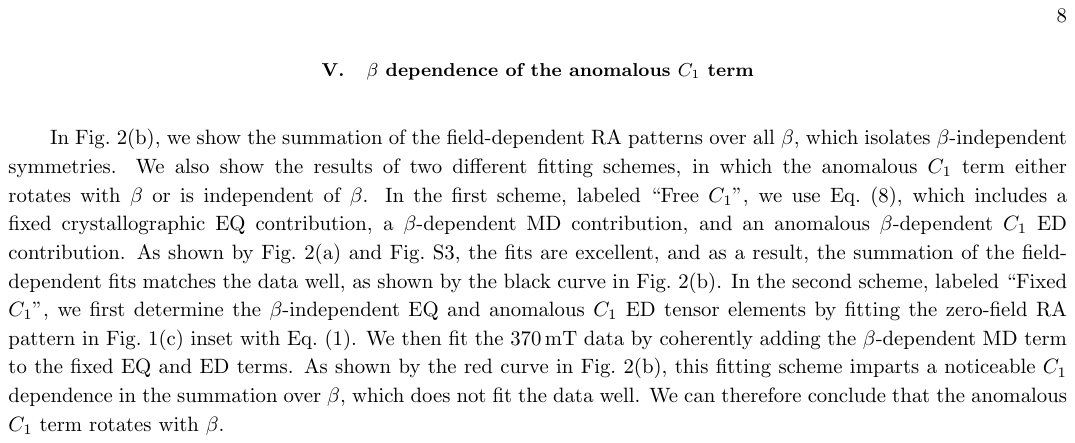}
    \caption{Caption}
    \label{fig:my_label}
\end{figure}\begin{figure}
   \vspace*{-2cm}
   \hspace*{-2cm}
    \centering
    \includegraphics[page=9]{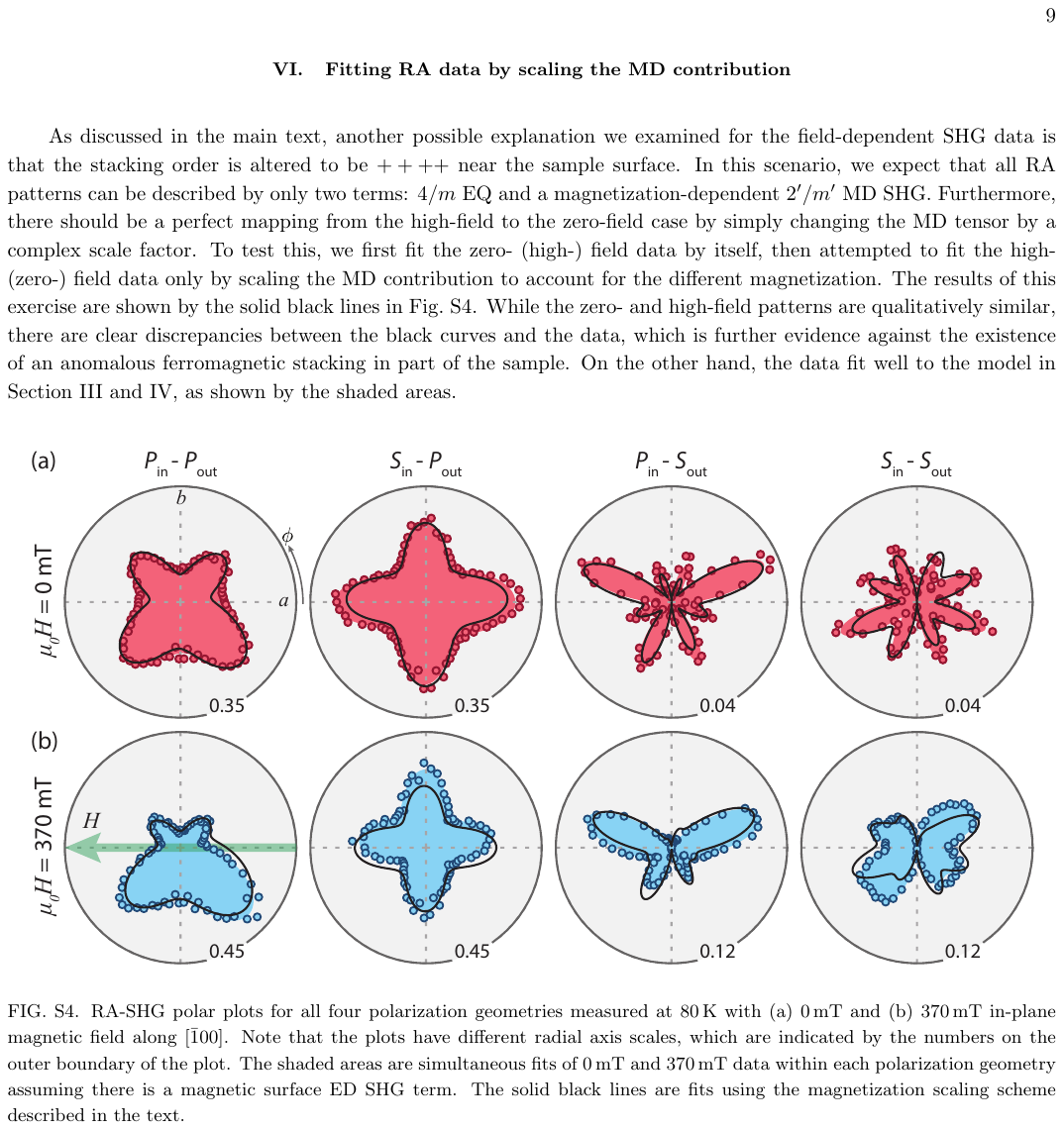}
    \caption{Caption}
    \label{fig:my_label}
\end{figure}
\begin{figure}
   \vspace*{-2cm}
   \hspace*{-2cm}
    \centering
    \includegraphics[page=10]{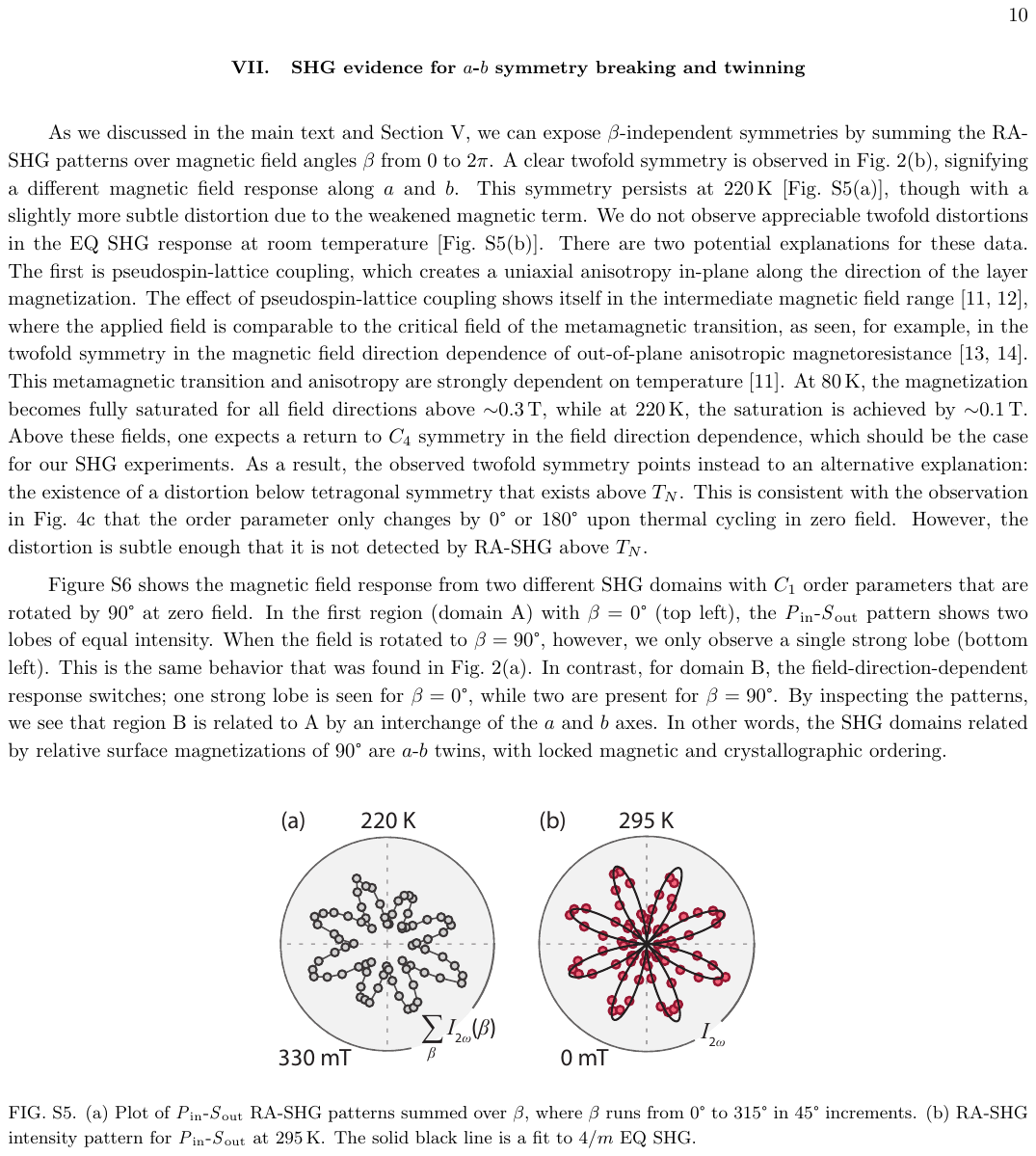}
    \caption{Caption}
    \label{fig:my_label}
\end{figure}
\begin{figure}
   \vspace*{-2cm}
   \hspace*{-2cm}
    \centering
    \includegraphics[page=11]{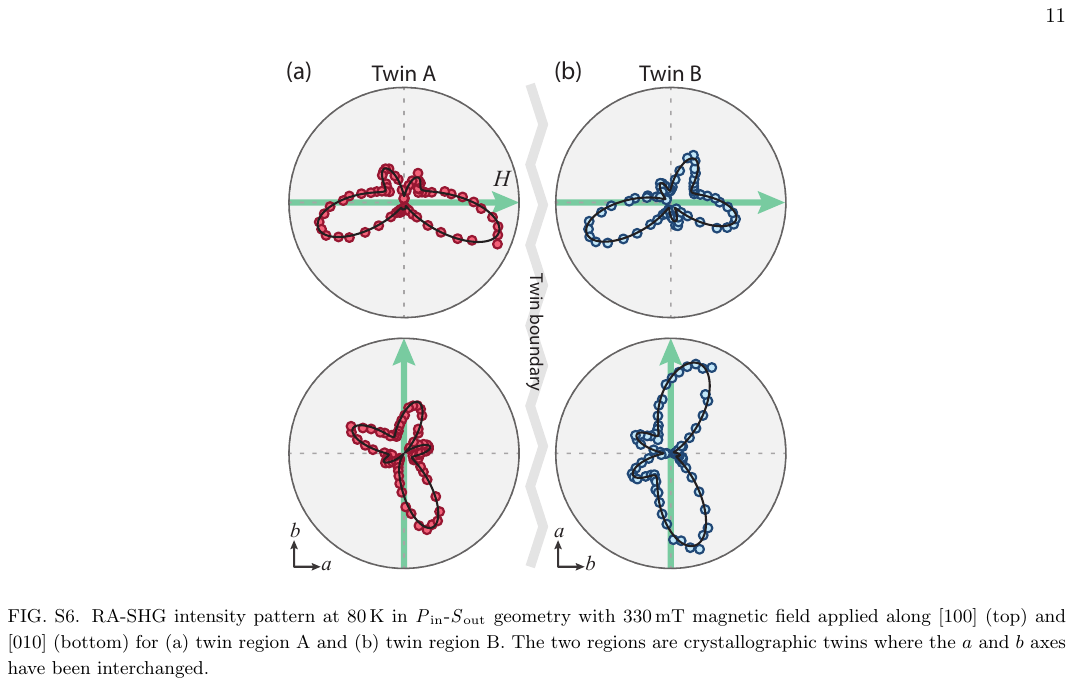}
    \caption{Caption}
    \label{fig:my_label}
\end{figure}
\begin{figure}
   \vspace*{-2cm}
   \hspace*{-2cm}
    \centering
    \includegraphics[page=12]{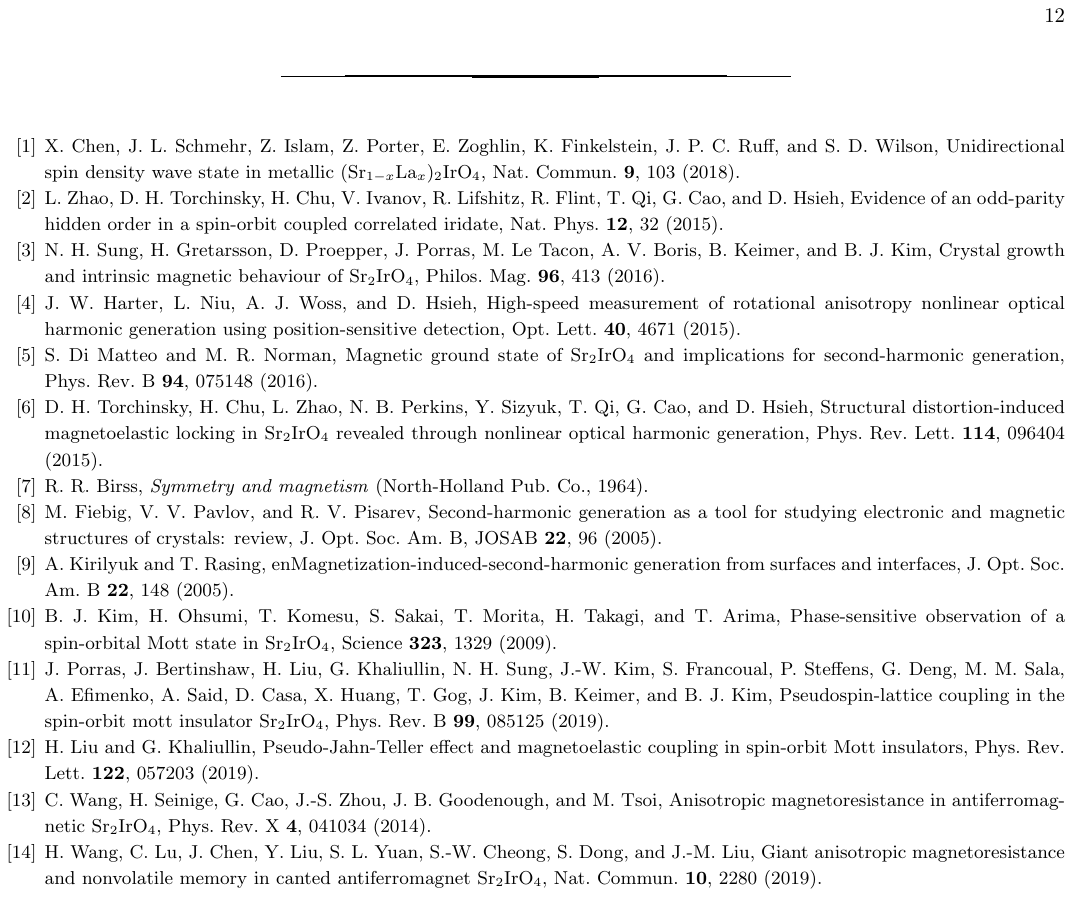}
    \caption{Caption}
    \label{fig:my_label}
\end{figure}
\end{document}